\documentclass[%
 reprint,
amsmath,amssymb,
aps,
prb,
showkeys
]{revtex4-2}

\usepackage{graphicx}
\usepackage{dcolumn}
\usepackage{bm}
\usepackage{hyperref}
\usepackage{blindtext}
\usepackage{siunitx}
\DeclareSIQualifier\rms{rms}
\usepackage{acronym}
\usepackage[noprefix]{nomencl}

\begin{document}


               
\newacro{AWG}{arbitrary waveform generator}
\newacro{FWHM}{full width at half maximum}
\newacro{LLG}{Landau-Lifshitz-Gilbert}
\newacro{LTEM}{Lorentz transmission electron microscopy}
\newacro{RF}{radio-frequency}
\newacro{STT}{spin-transfer torque}
\newacro{TEM}{transmission electron microscopy}
\newacro{UTEM}{ultrafast transmission electron microscopy}

\newcommand{\nomunit}[1]{%
\renewcommand{\nomentryend}{\hspace*{\fill}#1}}

\makenomenclature

\newcommand{\Sc}{c}
\nomenclature[c]{$\Sc$}{curl of the vortex}

\newcommand{\Sp}{p}
\nomenclature[p]{$\Sp$}{polarity of the vortex core}

\newcommand{\Sx}{x}
\newcommand{\Sy}{y}
\newcommand{\Sz}{z}
\nomenclature[x]{$\Sx,\Sy,\Sz$}{axes of the coordinate system}

\newcommand{\Snormvec}{\mathbf{e}}
\nomenclature[e]{$\Snormvec_i$}{uni vector along $i$}

\newcommand{\Sfex}{f_\text{ex}}
\nomenclature[fex]{$\Sfex$}{external driving frequency \nomunit{\si{\hertz}}}

\newcommand{\Sfzero}{f_0}
\nomenclature[f0]{$\Sfzero$}{free oscillation frequency \nomunit{\si{\hertz}}}

\newcommand{\Sfr}{f_\text{r}}
\nomenclature[fR]{\Sfr}{resonace frequency \nomunit{\si{\hertz}}}

\nomenclature[O.]{$\Omega$}{angular velocity of external driving frequency \nomunit{\si{\radian\per\second}}}

\nomenclature[G.]{$\Gamma$}{exponential damping parameter \nomunit{\si{\per\second}}}

\nomenclature[o.]{$\omega$}{angular velocity of free oscillation frequency \nomunit{\si{\radian\per\second}}}

\nomenclature[x.]{$\xi$}{non-adiabaticity parameter}

\newcommand{\Sj}{j}
\nomenclature[j]{$\Sj$}{electrical current density \nomunit{\si{\ampere\per\meter\squared}}}

\newcommand{\Is}{I_\text{S}}
\nomenclature[I]{$\Is$}{current through the sample \nomunit{\si{\ampere\rms}}}

\newcommand{\Msat}{M_\text{S}}
\nomenclature[MS]{$\Msat$}{saturation magnetization \nomunit{\si{\ampere\per\meter}}}

\nomenclature[t]{$t$}{time delay \nomunit{\si{\second}}}

\newcommand{\vC}{v_\text{C}}
\nomenclature[vC]{$\vC$}{velocity of the vortex core \nomunit{\si{\meter\per\second}}}

\newcommand{\frep}{f_\text{rep}}
\nomenclature[frep]{$\frep$}{\nomunit{\si{\hertz}}}

\nomenclature[D]{$D$}{vortex core orbit diameter \nomunit{\si{\meter}}}

\nomenclature[X,Y]{$X,Y$}{position coordinates of the vortex core \nomunit{\si{\meter}}}

\nomenclature[A]{$A$}{isotropic dilation factor}

\newcommand{\rot}{\hat{R}}
\nomenclature[R]{$\rot$}{rotation matrix}

\newcommand{\tx}{\alpha_\Sx}
\newcommand{\ty}{\alpha_\Sy}
\newcommand{\txy}{\alpha_{\Sx,\Sy}}
\nomenclature[alphax,alphay]{$\tx,\ty$}{tilt along $\Sx$ and $\Sy$ axis \nomunit{\si{\degree}}}

\newcommand{\Hfield}{H}
\newcommand{\Hfieldx}{H_\Sx}
\newcommand{\Hfieldy}{H_\Sy}
\newcommand{\Hfieldz}{H_\Sz}
\nomenclature[H]{$\Hfield$}{magnetic field strength \nomunit{\si{\ampere\per\meter}}}

\newcommand{\Erigid}{E_\text{quad}}
\nomenclature[Er]{$\Erigid$}{Energy of the vortex within the rigid vortex model \nomunit{\si{\electronvolt}}}

\newcommand{\Epin}{E_\text{pin}}
\nomenclature[Ep]{$\Epin$}{Depth of the pinning potential at a single pinning site \nomunit{\si{\electronvolt}}}

\newcommand{\Epinall}{E_\text{pin,all}}
\nomenclature[Epa]{$\Epinall$}{Sum of the individual pinning potentials \nomunit{\si{\electronvolt}}}

\newcommand{\Etotal}{E_\text{total}}
\nomenclature[Et]{$\Etotal$}{Total vortex potential, i.e. $\Etotal = \Erigid + \Epinall$ \nomunit{\si{\electronvolt}}}

\newcommand{\sigmapin}{\sigma_\text{pin}}
\nomenclature[sigma]{$\sigmapin$}{Width of a single pinning potential \nomunit{\si{\metre}}}

\newcommand{\sus}{\chi}
\nomenclature[chi]{$\sus$}{vortex displacement susceptibility \nomunit{\si{\meter\squared\per\ampere}}}

\newcommand{\susmod}{\tilde\chi}
\nomenclature[chimod]{$\susmod$}{modified displacement susceptibility $\modsus = \Hfield\sus$ \nomunit{\si[per-mode=symbol]{\meter\per\degree}}}

\newcommand{\stiff}{k}
\nomenclature[k]{$\stiff$}{stiffness factor of the restoring potential in the rigid vortex model \nomunit{\electronvolt\per\metre\squared}}

\nomenclature[rmed]{$r_\text{med}$}{median radial deviation between exerimental and simulated TRaPS dataset}

\title{Pinning and gyration dynamics of magnetic vortices\\
revealed by correlative Lorentz and bright-field imaging}


\author{Marcel Möller}
\email{marcel.moeller@uni-goettingen.de}
\author{John H. Gaida}
\author{Claus Ropers}
\email{claus.ropers@mpibpc.mpg.de}
\affiliation{
Max Planck Institute for Biophysical Chemistry, 37077 Göttingen, Germany\\
4th Physical Institute-Solids and Nanostructures, University of Göttingen, 37077 Göttingen, Germany
}

\date{\today}

\begin{abstract}
Topological magnetic textures are of great interest in various scientific and technological fields.
To allow for precise control of nanoscale magnetism, it is of great importance to understand the role of intrinsic defects in the host material.
Here, we use conventional and time-resolved Lorentz microscopy to study the effect of grain size in polycrystalline permalloy films on the pinning and gyration orbits of vortex cores inside magnetic nanoislands.
To assess static pinning, we use in-plane magnetic fields to shift the core across the island while recording its position. This enables us to produce highly accurate two-dimensional maps of pinning sites.
Based on this technique, we can generate a quantitative map of the pinning potential for the core, which we identify as being governed by grain boundaries. Furthermore, we investigate the effects of pinning on the dynamic behavior of the vortex core using stroboscopic Lorentz microscopy, harnessing a new photoemission source that accelerates image acquisition by about two orders of magnitude. We find characteristic changes to the vortex gyration in the form of increased dissipation and enhanced bistability in samples with larger grains.
 
\end{abstract}

\maketitle

\section{\label{sec:introduction}Introduction}

Microscopic magnetic objects such as vortices~\cite{Hubert1998,Okuno2000,Eggebrecht2017} and skyrmions~\cite{Skyrme1962,Muhlbauer2009,Yu2010} have attracted a sustained interest in the past decade.
Due to their stability and unique topological properties, these textures have sparked ideas for a vast number of technological applications such as (racetrack) memories~\cite{Parkin2008,Kiselev2011,Fert2013,Fert2017}, logical-gates~\cite{Jung2012,Zhang2015} and neuromorphic computing~\cite{Torrejon2017}.
While various control schemes for the manipulation of these textures by means of external magnetic fields~\cite{Choe2004,VanWaeyenberge2006}, electrical currents~\cite{Yamada2007,Woo2016,Gerlinger2021}, or optical pulses~\cite{Eggebrecht2017,Je2018,Gerlinger2021} are widely established, studies probing their interaction with defects are still highly sought after.

Previous investigations addressed the influence of artificial~\cite{Rahm2004,Uhlig2005,Vansteenkiste2008,Holl2020} and intrinsic~\cite{Burgess2013c,Compton2006,Compton2010,Chen2012,Chen2012b,Badea2016,Badea2016a} defects on the static pinning~\cite{Rahm2004,Uhlig2005,Burgess2013c,Holl2020} and the dynamics of magnetic vortices~\cite{Compton2006,Vansteenkiste2008,Compton2010,Kim2010c,Chen2012,Chen2012b}.
A majority of these studies utilized magneto-optical microscopy~\cite{Compton2006,Compton2010,Chen2012,Chen2012b,Badea2016}, which helped, for example, to link pinning to surface roughness in soft magnetic samples~\cite{Chen2012}.
Higher real-space resolution is offered by spin-polarized scanning tunneling microscopy in scenarios with atomic-scale defects on flat surfaces~\cite{Hanneken2016,Holl2020}.
This method revealed a Sombrero-shaped pinning potential between vortex cores and surface adsorbates~\cite{Holl2020}. 
Excellent spatial resolution for polycrystalline samples with higher surface roughness is possible using electron microscopy techniques~\cite{McVitie2015} and was used to characterize core pinning via one-dimensional differential-phase-contrast line scans~\cite{Uhlig2005}.
Moreover, recent advances in time-resolved Lorentz microscopy enable imaging magnetic dynamics at simultaneous high spatial and temporal resolution~\cite{Schliep2017,DaSilva2017,Zhang2019,Moller2020,Fu2020c,Cao2021}.
To date, however, the full capabilities of electron microscopy in both magnetic and structural imaging have yet to be leveraged in correlated studies.

In this work, we investigate the influence of the microcrystalline structure of permalloy thin films on the pinning of vortex cores by correlating the grain structure in bright-field images to the magnetic configuration in Lorentz micrographs.
In order to obtain maps of pinning sites, we developed TRaPS (\textbf{T}EM \textbf{Ra}stering of \textbf{P}inning \textbf{S}ites). This procedure locates defects by laterally shifting the vortex core across a nanostructure and imaging its position with high resolution. This allows us to directly calculate a quantitative two-dimensional representation of the pinning potential.
Moreover, we use time-resolved Lorentz microscopy with an improved photoemission source to assess the effect of grain size on the core gyration.
We find that annealing leads to a more corrugated pinning potential and larger average distances between pinning sites. Our findings suggest preferential pinning at grain boundaries and vortex orbits that avoid particularly large grains, demonstrating the combined strengths of correlated and \textit{in-situ} magnetic and structural characterization.
\newpage

\section{\label{sec:methods}Methods}

\subsection{\label{sec:sample}Sample System}

The sample system we investigate is a magnetic vortex confined in a square permalloy (Ni$_{80}$Fe$_{20}$) nanoisland~\cite{Okuno2000,Hubert1998}.
A magnetic vortex is a flux-closure type domain configuration which is predominantly oriented in-plane and curls either counter-clockwise ($c = +1$) or clockwise ($c = -1$) around its central core.
At the core region, the magnetization rotates out-of-plane and, either points up or down, said to have a polarization of $p=+1$ or $p=-1$, respectively~\cite{Okuno2000}.

A schematic representation of the sample is depicted in Figs.~\ref{fig:sketchExperiment}~a,b (light and scanning-electron images of the sample are in Supp. Fig.~4).
The permalloy square has a thickness of \SI{30}{\nano\meter} and an edge length of \SI{2}{\micro\meter}.
To electronically excite the sample, we overlap \SI{100}{\nano\meter} thick gold contacts on two opposing sides of the square, which extend to wire-bonding pads.
The nanoisland is positioned at the center of a \SI{15x15}{\micro\meter} large amorphous silicon nitride window.
At a thickness of \SI{30}{\nano\meter} the window is near electron-transparent and is supported by a \SI{200}{\micro\meter} thick silicon frame.

The sample fabrication processes involve electron-beam lithography using a positive-tone electron resist as well as electron-beam and thermal evaporation for the deposition of permalloy and gold, respectively.
We take special care to remove any resist prior to metal deposition by subjecting the developed sample to a short oxygen plasma.

The contacts on either side of the microstructure allow us to excite the vortex with in-plane \acs{RF}-currents, forcing its core on an elliptical trajectory~\cite{Kasai2006,Bolte2008,Pollard2012}.
This gyrotropic motion is a consequence of a combination of spin-transfer torques and current-induced Oersted fields~\cite{Kruger2007,Bolte2008}.
Generally, these systems allow for a resonance frequency between \SI{10}{\mega\hertz} and \SI{10}{GHz}, depending on the material and nanostructure size.
For the parameters of the current sample, we expect resonance frequencies around \SI{100}{\mega\hertz}~\cite{Moller2020}.

To identify the effects of the nanocrystalline structure on the pinning of vortex core, we investigate one annealed sample \textbf{A} as well as two non-annealed samples \textbf{N1} and \textbf{N2}.
We prepared all three samples under identical conditions on the same silicon frame.
Annealing is carried out by heating a single nanostructure using a low-frequency alternating current and monitoring the progress via a resistance measurement (for further details, see Suppl. Note 2).

\subsection{\label{sec:traps}TRaPS}

\begin{figure}
    \includegraphics[width=8.6cm]{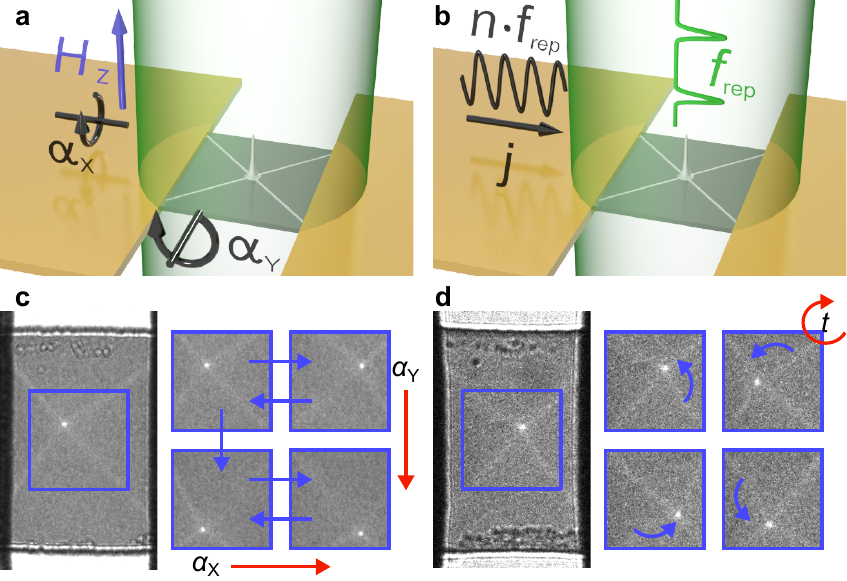}
\caption{
    \label{fig:sketchExperiment}
    (a) Sketch of the TRaPS measurement: The magnetic nanoisland (dark grey) is illuminated with a continuous collimated electron beam (green) and subjected to an external magnetic field $\Hfieldz$.
    Tilting the sample along the two tilt axes $\tx$ and $\ty$ enables us to move the vortex core within the nanostructure.
    (b) Sketch of the time-resolved experimental setup: Here, the nanoisland is imaged using a pulsed electron beam at a repetition rate of $\frep$ and excited with a synchronized alternating current at an integer multiple of the frequency of $\Sfex = n\cdot \frep$.
    (c) Four micrographs illustrating a TRaPS measurement. The vortex core is scanned horizontally back and forth across the island using $\tx$, while $\ty$ is gradually increased to vertically offset the core in between scans, as is illustrated by the blue arrows.
    (d) Example of four time-resolved Lorentz-micrographs acquired at different time delays between exciting current and the pulsed electron beam.
    The bright lines indicate the position of the domain walls within the nanostructure, with their intersection marking the location of the vortex core.
    Between frames, the core is visibly displaced and moving on a counter-clockwise trajectory.
}
\end{figure}

We introduce ``\textbf{T}EM \textbf{Ra}stering of \textbf{P}inning \textbf{S}ites'' (TRaPS) as a method to identify and locate static magnetic pinning sites using Fresnel-mode Lorentz microscopy~\cite{DeGraef2001,Zweck2016}.
For our sample, this imaging method visually highlights the position of the four domain walls as bright lines (cf. Fig.~\ref{fig:sketchExperiment}~c).
At their intersection, a peak is formed, marking the position of the vortex core.
To perform a TRaPS measurement, we shift the core across the film in a rasterized fashion using in-plane magnetic fields and record Lorentz micrographs at each raster step.
A repeated occurrence of the core at the same position can then indicate the location of a pinning site.

We generate the in-plane magnetic field components $\Hfieldx$, $\Hfieldy$ along the $\Sx$ and $\Sy$ directions of the sample by applying an out-of-plane field $\Hfieldz$ and tilting the sample along the two tilt axes $\tx$,$\ty$, as indicated in Figs.~\ref{fig:sketchExperiment}~a,c.
For sufficiently small angles, the resulting field components are $(\Hfieldx,\Hfieldy) \approx \Hfieldz \cdot (\ty, \tx)$.
The out-of-plane magnetic field $\Hfieldz$ is generated by weakly exciting the main objective lens of the microscope while using the objective mini-lens for imaging. 

\subsection{\label{sec:UTEM}Time-resolved Lorentz microscopy}

Time-resolved Lorentz microscopy is carried out at the Göttingen ultrafast transmission electron microscope (\acs{UTEM}\acused{UTEM}), a modified JEOL 2100F transmission electron microscope (\acs{TEM}\acused{TEM}) featuring a high-brightness photoemission electron source~\cite{Feist2017,Moller2020}.
For the present study, we equipped the \acs{UTEM} with a \ac{RF}-generator that is electronically synchronized to the photoemission laser, using a methodology introduced in Ref.~\cite{Moller2020}.
One output of the \ac{RF}-generator (Keysight 81160A) feeds into a custom \ac{TEM} holder that enables \ac{TEM} imaging under \textit{in-situ} current excitation with frequencies up to the GHz regime.
A second output triggers the photoemission laser, a gain-switched diode laser (custom Onefive Katana 05-HP) operating at a wavelength of \SI{532}{\nano\meter}. 
With its continuously variable repetition rate of $\frep= \SIrange{20}{80}{\mega\hertz}$ and a pulse duration of \SI{35}{\pico\second}, this laser enables us to increase the electron-pulse duty cycle by more than two orders of magnitude compared to previous studies~\cite{Moller2020}.
Thus, we can reduce the image acquisition time of a single stroboscopic micrograph from several minutes to a few seconds and compile much larger data sets that allow deeper insights into the vortex gyration.

We compile time-resolved movies of the vortex dynamics by exciting it at frequencies that are integer multiples of the laser repetition rate $\Sfex = n\cdot \frep$ and by incrementally changing the excitation phase between frames (see Fig.~\ref{fig:sketchExperiment}~b,d). 

\subsection{Bright-field imaging and vortex core localization}

To assess the nanocrystalline structure of our samples, we recorded bright-field images in low-magnification mode (Figs.~\ref{fig:OverviewAsdepositedAnnealed}~a,b) by filtering out scattered parts of the beam using an aperture.
This allows us to compare the vortex core positions we find in time-resolved and TRaPS measurements to the grain structure of the films.
Therefore, we track the core in the Lorentz micrographs and map the results on top of the bright-field images.
The tracking process involves calculating the center-of-mass of the pixels corresponding to the bright peak at the core position, which we identify as the largest pixel cluster above an intensity threshold.
This method enables a core localization with few-nanometer precision~\cite{Moller2020}.
To map the core positions onto the bright-field images, we use a geometric transformation~\cite{Goshtasby1988}, which is derived from the location of easily identifiable image features in both Lorentz and bright-field images.
Using this approach, we can translate position information between both reference frames.

\section{\label{sec:results}Results}

\subsection{TRaPS Measurements}
\label{sec:resultsTRaPS}

\begin{figure}[t]
    \includegraphics[width=8.6cm]{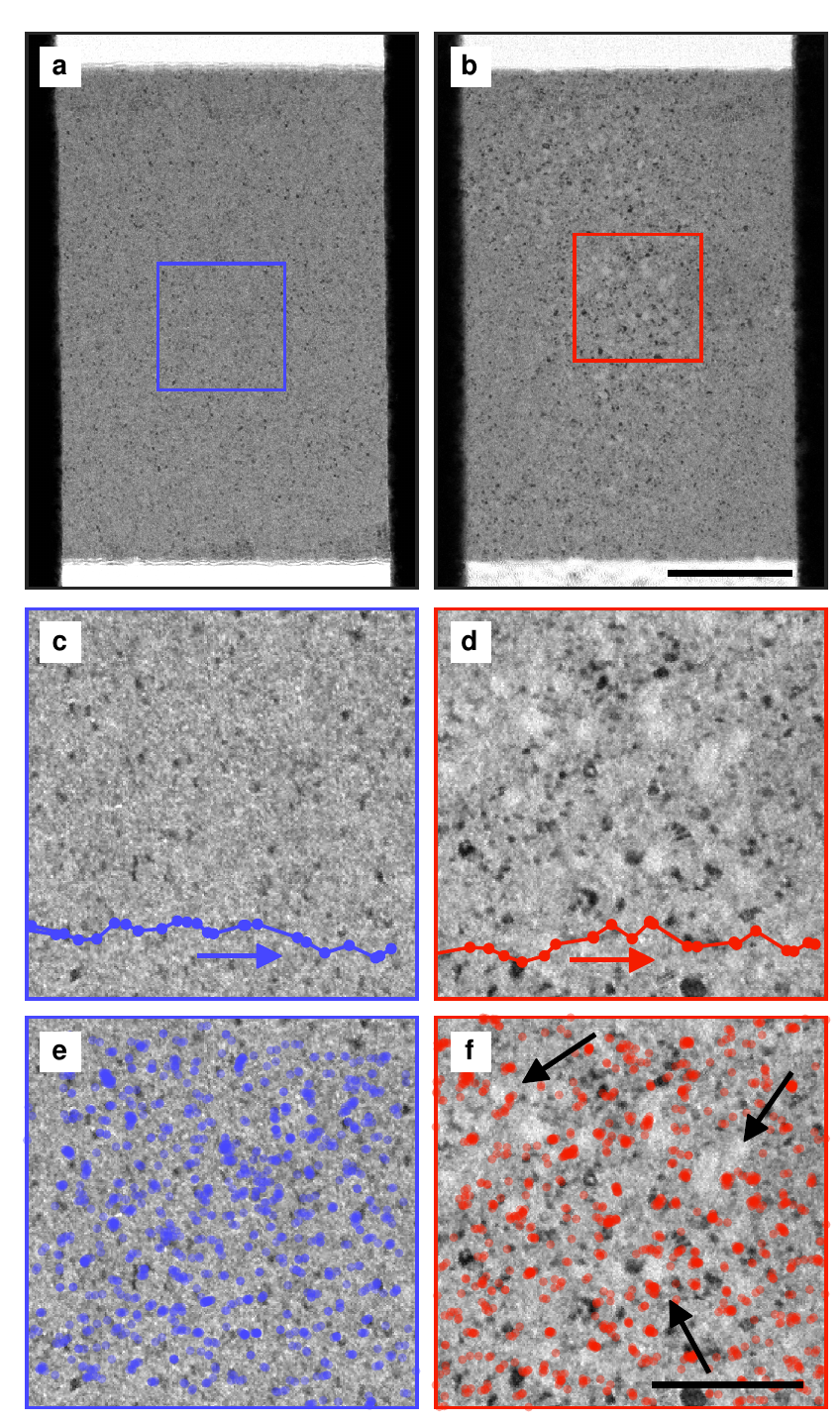}
    \caption{\label{fig:OverviewAsdepositedAnnealed}
        Bright field images of the non-annealed sample \textbf{N1} (a) and the annealed sample \textbf{A} (b, \SI{500}{\nano\metre} scalebar).
        (c,d) Close-ups of the regions marked in (a) and (b). The annealed sample shows a drastically increased grain size in the depicted area.
        The lines indicate the path of the vortex core during an exemplary tilt of primary tilt axis $\tx$, while $\ty$ is kept constant.
        (d,e) Tracked vortex core positions overlaid on top of the bright-field images in (c) and (d) (\SI{200}{\nano\metre} scalebar). 
    }
\end{figure}

TRaPS measurements are performed on the non-annealed sample \textbf{N1} and on sample \textbf{A}.
In their bright-field images (Figs.~\ref{fig:OverviewAsdepositedAnnealed}~a,b), the bright and dark regions surrounding the grey permalloy nanostructure are bare silicon-nitride and the opaque gold contacts, respectively.
A thin residue of permalloy, stemming from a partially coated undercut of the electron beam resist, is faintly visible at the top and bottom of the squares and can also be seen in the Lorentz images (e.g. Fig.~\ref{fig:sketchExperiment}~d).
Contrast variations in the films arise from spatially varying diffraction conditions of differently oriented grains and reveal a significant increase in grain size in the annealed sample (cf. Figs.~\ref{fig:OverviewAsdepositedAnnealed}~c,d).
This change in grain size is less significant in the vicinity of the gold contacts as these locally increase the thermal coupling, resulting in an inhomogeneous temperature profile during the annealing process.

For the TRaPS measurements, we apply an external field of $\Hfieldz =  \SI{35.8}{\kilo\ampere\per\meter}$ and vary both tilt axes in a range from \SI{-2.2}{\degree} to \SI{2.2}{\degree} in increments of \SI{0.2}{\degree}.
Along the primary tilt direction $\tx$ the samples are tilted back and forth once for every tilt position of the secondary axis $\ty$.
Additionally, each sweep of $\tx$ includes an approach step at either $\tx = \SI{3.0}{\degree}$ or \SI{-3.0}{\degree}.
Supplementary Movies 1 and 2 show the complete TRaPS measurements of either sample, with some example micrographs of sample \textbf{N1} presented in Fig.~\ref{fig:sketchExperiment}~c.

As an example, we marked the path of the core during one sweep of $\tx$ in Figs.~\ref{fig:OverviewAsdepositedAnnealed}~c,d, where we see that the core does not move in straight lines but rather zigzags and occasionally gets trapped.
All tracked positions are marked with dots in Figs.~\ref{fig:OverviewAsdepositedAnnealed}~e,f where they are plotted on top of the bright-field image of the corresponding region.
Due to the small size of the grains in sample \textbf{N1}, there is no apparent visual correspondence between the nanocrystalline structure and the core positions (Fig.~\ref{fig:OverviewAsdepositedAnnealed}~e).
However, for the annealed sample \textbf{A}, we find numerous pinning sites located directly at boundaries of larger grains (see arrows Fig.~\ref{fig:OverviewAsdepositedAnnealed}~f).
Furthermore, it stands out that the core never resides within one of the large grains.
Both observations clearly demonstrate that grain boundaries can pin vortex cores in polycrystalline films.
While this behavior was suspected before~\cite{Badea2016}, it has, to our knowledge, never been directly observed.

\begin{figure}[h]
\includegraphics{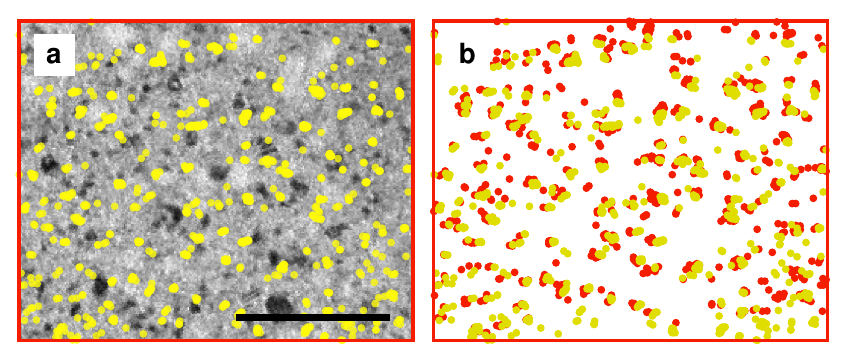}
\caption{\label{fig:AnnealedComparisonBothTilts}
    Comparison between TRaPS measurement on sample \textbf{A} with clockwise curl $\Sc = -1$ and underfocus imaging conditions (red dots) and counter-clockwise curl $\Sc = +1$ and overfocus imaging conditions (yellow dots).
    (a) Tracked vortex core position overlaid onto bright-field image of sample \textbf{A} (\SI{200}{\nano\metre} scalebar).
    (b) Depiction of the tracked vortex position of both measurements in the reference frame of the bright-field image in (a).
    Both datasets are in good agreement with each other, with only a minor offset between the two.
    The difference in the overall center of both TRaPS measurements can be explained with a residual magnetic in-plane field at the sample position (see text).}
\end{figure}

To assess the accuracy of our measurement technique, we repeat the tracking and the transform calculation for a second set of experimental conditions on sample \textbf{A}.
The results presented in Fig.~\ref{fig:OverviewAsdepositedAnnealed}~d are acquired with a clockwise curl ($\Sc=-1$), necessitating underfocused imaging conditions to achieve a bright spot at the core position~\cite{Schneider2000}.
For the second measurement, we altered the domain state to a counter-clockwise curl ($ \Sc = +1$), and acquire overfocused Lorentz images (see Suppl. Movie 3).
This second set of tracking data was likewise correlated with the bright-field image as shown in Fig.~\ref{fig:AnnealedComparisonBothTilts}~a, and is in excellent agreement with the previous results.
This is particularly evident when we plot both data sets together in the same reference frame as in Fig.~\ref{fig:AnnealedComparisonBothTilts}~b, where the only discernible difference is a minor lateral displacement of about \SI{6}{\nano\meter}.
In addition to confirming that TRaPS allows for accurate localization of pinning sites, this comparison also demonstrates that the underlying process is independent of the vortex curl.

\begin{figure}
    \includegraphics{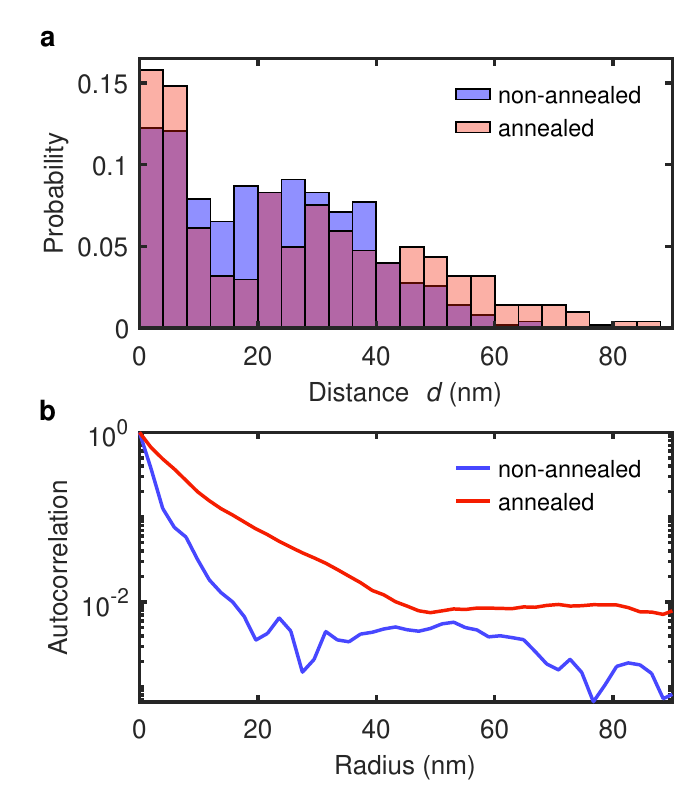}
    \caption{\label{fig:AutocorrelationDistancedistribution}
    (a) Probability that the vortex core moved by distance $d$ between two TRaPS steps. 
    (b) Radial autocorrelation of the bright field images displayed in Fig.~\ref{fig:AnnealedComparisonBothTilts}~c,d.
    }
\end{figure}

To further analyze our data, we calculate the distance $d$ traveled by the vortex between two consecutive steps of a TRaPS scan.
Fig.~\ref{fig:AutocorrelationDistancedistribution}~b shows histograms of the jump distances for both samples (the initialization steps at $\tx = \SI{\pm3}{\degree}$ are not included in these statistics).
The most prominent feature in both distributions is a strong peak at small distances $d < \SI{8}{\nano\metre}$.
For these distances, we can assume that the vortex has remained at the same pinning site, and as we would expect, this happens more frequently for sample \textbf{A}.
In contrast, we identify substantial differences between both samples for distances $d> \SI{8}{\nano\metre}$:
Firstly, the vortex in the annealed sample rarely moves by distances in the range of \SI{8}{\nano\meter} to \SI{20}{\nano\meter}, and secondly, it more frequently jumps over longer distances beyond \SI{40}{\nano\meter}, which evidently is a result of larger grain sizes.

For comparison, we also compute the average grain size in both samples using the full-width-half-maximum of the radial autocorrelation function~\cite{Heilbronner1992,Zang2018}\footnote{It is worth noting that dark-field images were used for this assessment in the original publications. However, we believe that for a rough estimate, we are interested here, bright-field images are equally suitable}.
From the autocorrelation functions in Fig.~\ref{fig:AutocorrelationDistancedistribution} we estimate grain sizes of \SI{4}{\nano\meter} and \SI{10}{\nano\meter} for sample \textbf{N1} and \textbf{A}, respectively.
This demonstrates that the typical jump distances span up to several grain diameters, suggesting that not every grain boundary causes effective pinning.

\subsection{\label{sec:PinningPotential}Quantifying the Pinning Potential}

\begin{figure}
    \includegraphics[width=\columnwidth]{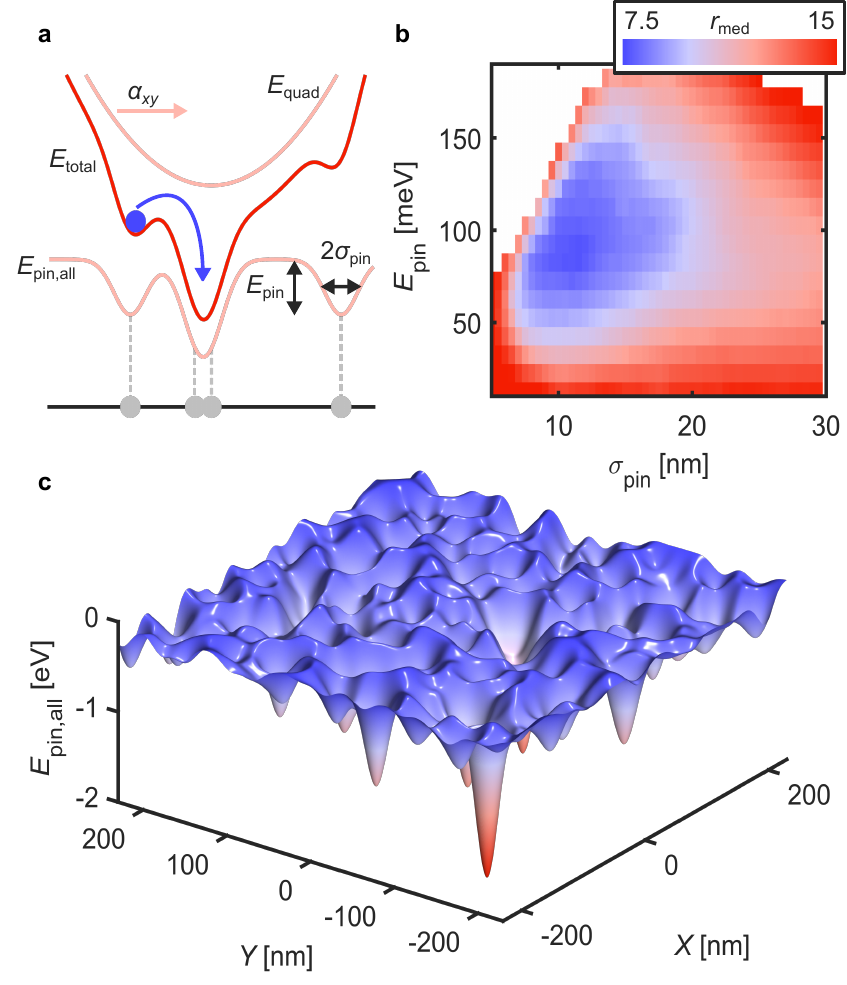}
    \caption{
    Reconstruction of the pinning potential from TRaPS data of sample \textbf{N1}.
    (a) Illustration of model potential $\Etotal$ used for the simulated TRaPS data.
    $\Erigid$ is a quadratic plus a field-dependent linear potential (here, as a function of the tilt angle $\txy$) that defines the equilibrium position of the vortex core in the absence of any pinning.
    $\Epinall$ is composed of a Gaussian potentials for each experimentally measured pinning position (grey dots), all having the same depth $\Epin$ and width $\sigmapin$.
    By increasing/decreasing the tilt angle, the vortex core is dragged along and probes the pinning potential.
    (b) Median radial deviation $r_\text{med}$ between the experimental and simulated TRaPS data for various combinations of $\Epin$ and width $\sigmapin$. 
    The best agreement is obtained at $r_\text{med} = \SI{7.6}{\nano\metre}$ with $\Epin = \SI{80}{\milli\electronvolt}$ and $\sigmapin = \SI{11}{\nano\metre}$.
    Values of $\Epin$ and $\sigmapin$ which do not improve $r_\text{med}$, compared to a case of no pinning potential, are left blank.
    (c) 2D representation of the pinning potential $\Epinall$ for $\Epin = \SI{80}{\milli\electronvolt}$ and $\sigmapin = \SI{11}{\nano\metre}$. 
    }
    \label{fig:PinningPotential}
\end{figure}

In order to extract the pinning energy landscape from our TRaPS measurement, we compare experimental with simulated TRaPS data.
Therefore, we model the movement of the vortex core in a global quadratic plus a local pinning potential, as illustrated in Fig.~\ref{fig:PinningPotential}~a.
By scanning the quadratic potential across the disorder potential, we replicate how the core is trapped and moves between pinning sites.

We define the quadratic potential using the rigid vortex model~\cite{Guslienko2001a}, for which the energy of the vortex domain configuration is expressed in terms of the core position $(X,Y)$ and the external magnetic field
\begin{align}
   \Erigid = \frac{1}{2} \stiff (X^2+Y^2) + c k \sus\left( -\Hfieldx Y + \Hfieldy X  \right)\;. 
\end{align}
Here, $\stiff$ is the stiffness factor of the quadratic potential, and $\sus$ is the displacement susceptibility.
For our sample geometry, we calculate a stiffness factor of $\stiff = \SI{1.63e-3}{\electronvolt\per\nano\metre\squared}$ using micromagnetic simulations~\cite{Vansteenkiste2014}.
The simulation is performed with a $512 \times 512 \times 4$ cell geometry, an exchange coupling of $A_\text{ex} = \SI{1.11e-11}{\joule\per\metre}$~\cite{Yin2015}, and a saturation magnetization of $\Msat = \SI{440}{\kilo\ampere\per\metre}$\footnote{This value was obtained from a comparison of experimental and simulated Lorentz images, using the nominal film thickness of \SI{30}{\nano\metre}. It is somewhat lower than expected from bulk values, which might be caused by uncertainty in the determination of the film thickness}. 

The equilibrium core position, i.e., the minimum of $\Erigid$, is given by
\begin{align}
    \begin{pmatrix}
        X \\ Y
    \end{pmatrix}
    = c \sus 
    \begin{pmatrix}
        - \Hfieldy \\ \Hfieldx    
    \end{pmatrix}
    = c \underbrace{\sus \Hfieldz}_{=: \susmod}
    \begin{pmatrix}
        -\tx \\ \ty
    \end{pmatrix}\;.
    \label{eq:susmod}
\end{align}
Here, we replace $\sus$ with a modified displacement susceptibility $\susmod = \sus \Hfieldz$, which specifies the core movement per tilt angle.
By fitting Eq.~\ref{eq:susmod} to our data, we find $\susmod$ to be \SI[per-mode=symbol]{98}{\nano\meter\per\degree} and \SI[per-mode=symbol]{108}{\nano\meter\per\degree} for sample \textbf{N1} and \textbf{A}, respectively.
This is in good agreement with the micromagnetic simulation, which predicted a value of $\susmod = \SI[per-mode=symbol]{123}{\nano\meter\per\degree}$.

To simulate the behavior of the pinning sites, we place a Gaussian-potential dip with depth $\Epin$ and width $\sigmapin$ at every position $\left(X_{i,\text{exp}},Y_{i,\text{exp}}\right)$ tracked in the TRaPS measurement (see Fig.~\ref{fig:PinningPotential}a). 
This ensures a deeper and/or broader potential in regions where the core is encountered more frequently.
The pinning potential is hence given by 
\begin{align}
    \Epinall = - \Epin \sum_i \exp \left( - \frac{r^2_i}{2 \sigmapin^2}\right)\;,
\end{align}
where $r^2_i = \left(X -X_{i,\text{exp}}\right)^2 + \left(Y - Y_{i,\text{exp}}\right)^2$ is the distance to the core position  measured at tilt step $i$ of the TRaPS measurement.
The total potential is thus $\Etotal = \Erigid + \Epinall$.

In the course of a single simulation we set the same sequence of tilt angles and, at each tilt step $i$, find the next local minimum $\left(X_{i,\text{sim}},Y_{i,\text{sim}}\right)$ of $\Etotal$ in the direction of steepest descent.
The starting point of this minimization is the simulated minimum from the previous tilt step $i-1$, just like in the experiment.
This leaves us with a set of simulated core positions from which we can calculate the median radial deviation
\begin{align}
    r_\text{med} = \text{median}_i \left(
        \begin{Vmatrix}
            X_{i,\text{sim}} - X_{i,\text{exp}}\\
            Y_{i,\text{sim}} - Y_{i,\text{exp}}
        \end{Vmatrix}
    \right)\;.
\end{align}
Lastly, we find the combination of $\Epin$ and $\sigmapin$ (the only free parameters in our model) that minimizes $r_\text{med}$ and thus best represents our experimental data.

Figure~\ref{fig:PinningPotential}~b shows the median radial deviation for all simulations based on the TRaPS measurement of sample \textbf{N1}.
It is minimum at $\Epin = \SI{80}{\milli\electronvolt}$ and $\sigmapin = \SI{11}{\nano\metre}$.
In case of sample \textbf{A}, we obtain similar values of $\Epin = \SI{90}{\milli\electronvolt}$ and $\sigmapin = \SI{12}{\nano\metre}$, which results in an increase of the integrated pinning potential of about \SI{30}{\percent}.
The simulated core positions are in good agreement with experimental data, as we present in Suppl. Fig.~7, together with images of the pinning potential.

A three-dimensional representation of $\Epinall$ of sample \textbf{N1} is given in Fig.~\ref{fig:PinningPotential}~c.
It has a roughness, estimated by the standard deviation of the potential, of \SI{193}{\milli\electronvolt} and measures \SI{-2.0}{\electronvolt} at its deepest point.
For the annealed sample, the respective values are \SI{281}{\milli\electronvolt} and \SI{-2.4}{\electronvolt}. This analysis demonstrates that polycrystalline samples with larger grains show an overall increase in their pinning potential, with an increased roughness and deeper minima. 
Furthermore, we note that the demonstrated method is capable of quantifying pinning potentials down to sub-\SI{100}{\milli\electronvolt} and spanning multiple orders of magnitude.

\subsection{\label{sec:timeResolvedTrajectories} Time-resolved Trajectories}

Besides studying the static interaction of the vortex core with pinning sites, we also probed how the dynamic vortex gyration is affected by the nanocrystalline configuration.
We therefore perform time-resolved measurements on the non-annealed sample \textbf{N2} and the annealed sample \textbf{A}, where we excite the permalloy square with an alternating current forcing the vortex core on an elliptical trajectory.
We recorded this motion in sample \textbf{N2} for excitation frequencies from $\Sfex = \SI{86}{\mega\hertz}$ to  \SI{99.5}{\mega\hertz} at a current density of $\Sj = \SI{3.6E10}{\ampere\rms\per\meter\squared}$ and in case of sample \textbf{A} for frequencies from \SI{72}{\mega\hertz} to \SI{84}{\mega\hertz} at $\Sj = \SI{8.2E10}{\ampere\rms\per\meter\squared}$.
The higher currents and lower frequencies in the case of sample \textbf{A} were only necessary after annealing, whereas before, we were able to observe core gyration at similar excitation parameters as for sample \textbf{N2}.
To measure and ensure a constant excitation current throughout the frequency range, we monitor the sample current $\Is$ using an oscilloscope (for details, see Suppl. Note 1).

\begin{figure}
    \includegraphics[width=\columnwidth]{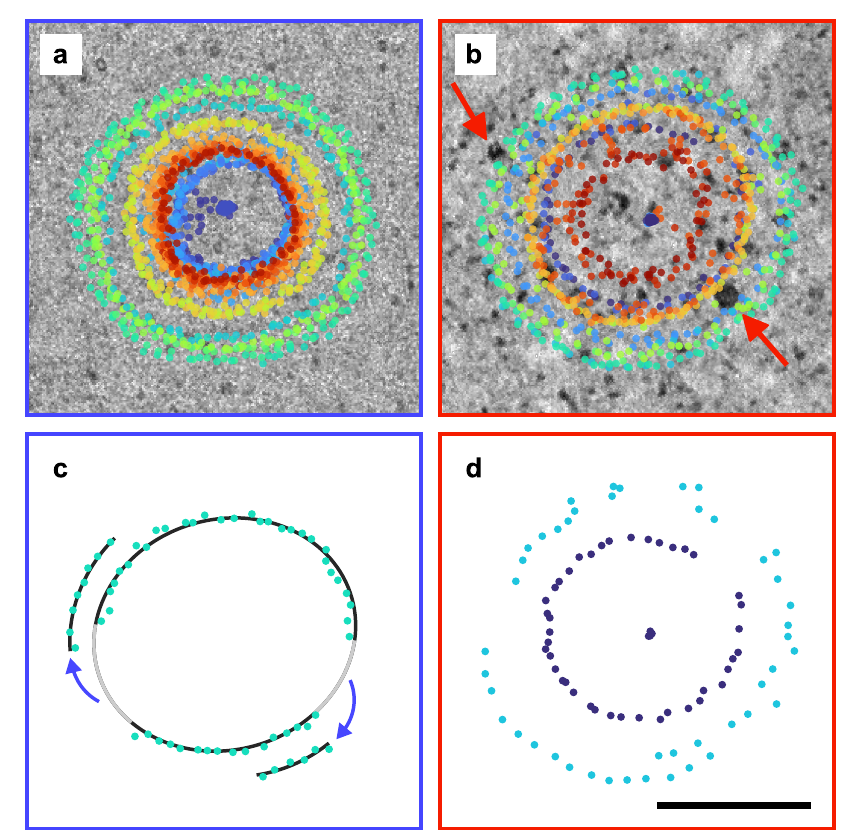}
    \caption{
        \label{fig:ComparisonTimeresolved}
        (a,b) Trajectories (dots) of the vortex gyration in sample \textbf{N2} (a) and sample \textbf{A} (b) measured via time-resolved Lorentz microscopy overlaid on top of bright-field images of samples. 
        Each color corresponds to a specific frequency according to the colorbar given in Fig.~\ref{fig:TimeResolvedReaction}.
        (c) Example of a segmented trajectory in (a) indicating a bistable gyration. The black line is a guide to the eye, indicating the movement of the core. The greyed-out path and the arrows indicate where the core jumped to another equally favorable orbit.
        (d) Two examples for jagged trajectories in (b) (\SI{200}{\nano\metre} scalebar).
        }
\end{figure}

At each frequency, we record up to 60 micrographs and incrementally increase the phase between the \ac{RF}-current and the probing electron beam to cover the whole excitation period.
These micrographs are combined into Supp. Movies~4 and 5, and show the time-resolved gyration together with the tracked core positions.
Figures~\ref{fig:ComparisonTimeresolved}~a,b show a compilation of all trajectories overlaid on the bright-field images (individual depictions in Suppl. Figs.~5 and 6).

\begin{figure}
    \includegraphics{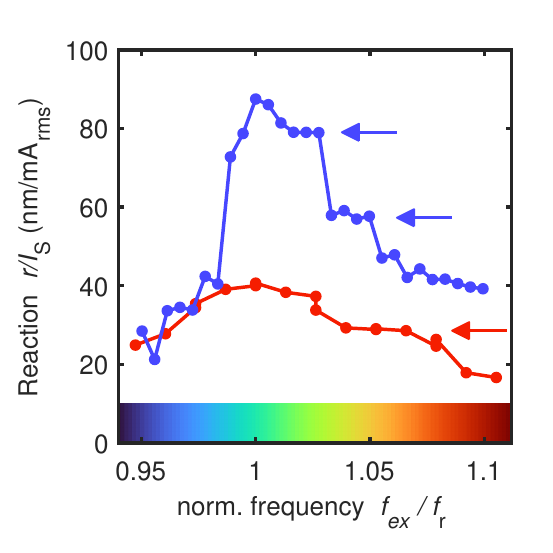}
    \caption{
        \label{fig:TimeResolvedReaction}
        Excitation-current-normalized radius $r/\Is$ as a function of normalized frequency $\Sfex/\Sfr$. For some frequency ranges the orbit radius stays constant upon changing the excitation frequency (arrows).
        }
\end{figure}

Similar to our TRaPS measurements, we can also identify grains in the annealed sample in Fig.~\ref{fig:ComparisonTimeresolved}~b that appear to be avoided by the core, however, less conclusively as in the static case.
We marked two of these grains with arrows in Fig.~\ref{fig:ComparisonTimeresolved}~b.
Furthermore, we find discontinuous jumps of the vortex position upon cycling the excitation phase, where it appears to switch between two or more equally favorable trajectories.
Interestingly, this behavior is encountered much more regularly for the annealed sample, both at a larger number of frequencies and during a measurement at a given frequency (see Supp. Figs.~5 and 6).
Most likely, the jumps between multistable orbits are triggered by the sudden (yet small) change of the excitation phase between time-resolved micrographs.
This multistability can be considered the dynamic counterpart of stochastic switching between bistable static pinning sites~\cite{Burgess2013c,Badea2016a,Badea2018}.

To further evaluate the trajectories, we fit them to an ellipse and determine the mean of both semiaxes $r$~\cite{Gal}.
Figure~\ref{fig:TimeResolvedReaction} shows the resulting radii $r$ diveded by the sample current $\Is$ and plotted against the normalized frequency $\Sfex/\Sfr$.
The graphs reveal a resonance frequency of the gyration of $\Sfr = \SI{90.5}{\mega\hertz}$ and \SI{76}{\mega\hertz} for sample \textbf{N2} and \textbf{A}, respectively.
Here, two observations stand out.
First, the radius of the trajectories does not change continuously in size as a function of frequency $\Sfex$ (see also Refs.~\cite{Pollard2012,Moller2020}), suggesting hysteretic behavior.
Instead, we find plateaus, for which trajectories cluster at certain radii (marked with arrows in Fig.~\ref{fig:TimeResolvedReaction}), clearly indicating an orbital stabilization by the pinning potential.
Secondly, we find that the current-normalized radius $r/\Is$ is significantly smaller and exhibits a broader resonance in the case of the annealed sample \textbf{A}, which is a distinct sign of enhanced dissipation in this sample. 
These two observations, together with enhanced multistability in the annealed sample, demonstrate that grain sizes have an important influence on the dynamic behavior of gyrating vortices.

\section{Conclusion\label{sec:conclusion}}

Nanocrystallinity and surface roughness have long been linked to the pinning of vortices in soft-magnetic films~\cite{Uhlig2005,Chen2012,Badea2016}. The direct real-space identification of grain boundaries as effective pinning sites for the core was enabled by the TRaPS method introduced in this study. The correlation of structural and magnetic imaging in electron microscopy can be further developed to trace the microscopic origins of pinning down to the atomic scale, combining high-resolution (scanning) TEM with holography~\cite{Lichte2008,Midgley2009} or differential phase contrast~\cite{Chapman1978,McVitie2015}.

The joint high spatial and temporal resolution of our approach will be critical to explore transient pinning and local damping effects, while the quantitative TRaPS potential will serve as input for future theoretical studies on driven vortex dynamics. The global ansatz for the calculation of the trapping potential based on common properties of pinning sites may be further refined by taking into consideration characteristics of individual defects. The observation of increased roughness, deeper traps and enhanced bistability for samples with larger grains may become relevant for device fabrication and in the tailoring of annealing processes to mitigate or selectively enhance pinning. 

Finally, on the methodical side, the two orders of magnitude increase in time-averaged brightness of the photoemission source will have immediate benefits in picosecond stroboscopic imaging of ultrafast dynamics also beyond magnetism, including nanoscale structural and electronic phenomena.

\begin{acknowledgments}
This work was funded by the Deutsche Forschungsgemeinschaft (DFG) in the Collaborative Research Center “Atomic Scale Control of Energy Conversion” (DFG-SFB 1073, project
A05) and via resources from the Gottfried Wilhelm Leibniz-Prize. We gratefully acknowledge support by the Lower Saxony Ministry of Science and Culture and funding of the instrumentation by the DFG and VolkswagenStiftung.
Furthermore, we acknowledge helpful discussions and assistance from the Göttingen UTEM Team, especially Thomas Danz, Till Domröse, Armin Feist and Karin Ahlborn. 
\end{acknowledgments}

C.R. and M.M. conceived the project. M.M. prepared the sample. M.M. and J.H.G. conducted the experiment. M.M. evaluated the data and discussed the results with J.H.G. and C.R. M.M. and C.R. wrote the paper with inputs from J.H.G.

\bibliography{library.bib}

\end{document}